# PAPG – Personalized Anti-Phishing Guard

**Belal Amro[1], Ahmed Abu Sabha[2], Ammar Qunaibi[3] & Ibraheem Najjar[4]**

*[1,2,3,4]Faculty of Information Technology, Hebron University, Hebron, Palestine*

**Abstract:** Security and privacy have been considered a corner stone in all electronic transactions nowadays. People are becoming very cautious when conducting electronic transactions over internet. One of the major issues that frightens them is identity theft. Identity theft might be conducted using phishing techniques that aims to trick the user to provide his credentials in a well-organized tactic. Efforts have been done towards fighting against phishing attacks and hence identify theft. However, most of these efforts are either computationally exhaustive to the electronic device or depend on a third party to perform the task. In this paper, we propose a plugin called Personalized Anti-Phishing Guard – PAPG that is managed personally on the device and is used to guard the user against phishing attacks. The plug-in maintains data locally and may not need to synchronize with a third party. Besides, PAPG depends on the user's feedback to build the local knowledge base that is used to support the decision. The user might also store his profile and reuse it with other devices and from different locations without having to configure it again.

## 1. INTRODUCTION

Identity theft is considered as one of the most important security challenges that needs to be well addressed and effectively flighted. According to Symantec internet security threat report, the number of URLs related to phishing activity has risen by more than 182% in the year 2017 that than of 2016[1]. Symantec defined phishing as "an attempt to illegally gather personal and financial information by sending a message that appears to be from a well-known and trusted company". The false message has a fraud link to a malicious webpage that is somehow very similar in design to the original one, the user is asked to provide his credentials to the fake webpage and hence his credentials will be disclosed. Phishing techniques have been deployed widely and spread over different services and technologies including mail service [2], web service[3], and mobile devices[4]. Referring to [1], The increase of phishing attacks in 2016 has grown over than 10% from the attacks identified during 2015.

Phishing techniques has been developing over the past years, according to [5], phishing is divided into two categories; namely deceptive phishing and malware based phishing. The first method mainly relies on social engineering to acquire the identity of the victim while the later depends on using some types of malicious software to perform the task.

In their 2018 security report [6], Symantec reported the percentage of infection phishing vectors as shown in Figure 1 depicted from their report. As seen from the figure, email phishing still the dominant method used till now.

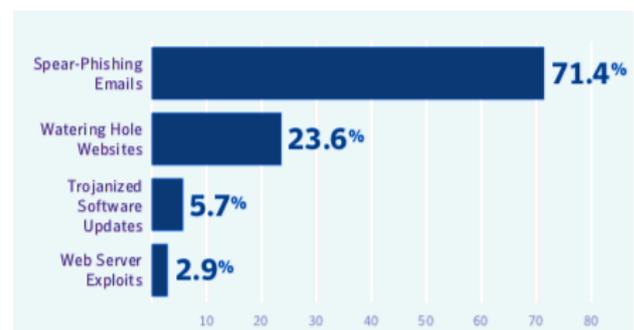

Figure 1: infection vectors of phishing attacks

In their survey about visual similarity based anti-phishing techniques, [13] provided a taxonomy of ant-phishing techniques. This taxonomy is shown in Figure 2.

*E-mail: bilala@hebron.edu, abusabha_ahmad@yahoo.com, abusabha_ahmad@yahoo.com, Ib1996rahim@hotmail.com*

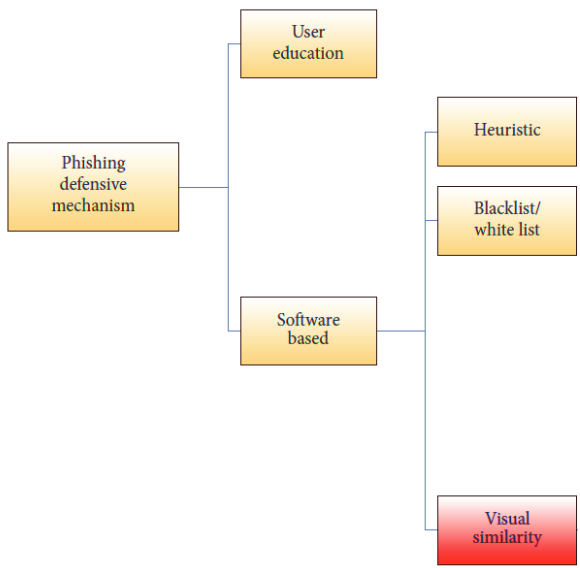

Figure 2: taxonomy of phishing techniques

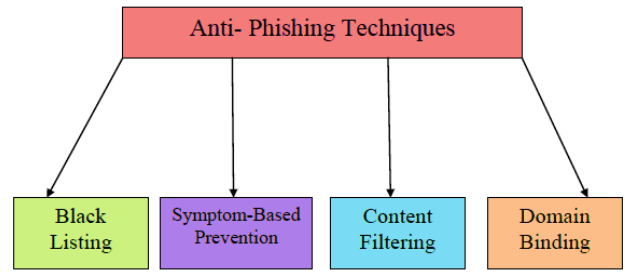

Figure 3: types of anti-phishing techniques

According to Figure 2. Phishing defense mechanisms are based on either user's education which aims at enhancing his knowledge about phishing attacks, or are based on software that automates the process.

We will focus here on website anti-phishing tools because all other methods including email phishing will lure the user to open a fake website for the gathering of credentials. Therefore, using a website anti-phishing tool might also be extended to reduce email phishing and other phishing attacks by testing email URLs against phishing websites.

The rest of the paper is organized as follows: in Section 2, we will provide a literature review of anti-phishing techniques. Section 3 will introduce our Personalized Anti-Phishing Guard (PAPG) plug-in. Performance analysis will be provided in Section 4. And Conclusion and Future work will be provided in Section 5.

## 2. Literature Review

Due to the widespread of phishing attacks and the losses associated with them, many anti-phishing techniques have been proposed to mitigate and reduce these attacks. According to [5], there are 4 categories of website anti-phishing techniques describes in Figure 2 below:

Black listing relies on comparing the given URL against a list of phishing URLs called black list [7][14][15][16]. The list can be updated from spam emails or from other sources such as Anti Phishing Working Group (APWG) and Phishtank. A tool called Netcraft is an anti-phishing solution that belong to blacklist based techniques. As discussed in Section 1, phishing websites are growing dramatically and hence the size of the blacklist might be huge which in turns may make this technique inefficient.

Symptom based anti-phishing prevention techniques analyze the content of a webpage and generate an alert according to the number and type of symptoms found [8]. Many of symptom based anti-phishing prevention techniques are based on visual similarity methods (document object model, visual features, CSS features,..,etc.). The phishing report is based on different attributes including block level similarity layout similarity and style similarity [22][23][24][25][26]. Visual similarity requires extensive calculations to be done; this makes it inefficient to be used by persons and preferred to be used by engines that deliver such services to its users.

Content filtering anti-phishing technique uses the content of the email or webpage to decide whether the content is phishing or not according to some Bayesian statistics and support vector machines SVM[9][17][18][19]. This method relies on artificial intelligence methods that also requires extensive calculations.

Domain binding is based on a client browser based technique where sensitive information is bound to a particular domain and a user is notified of which information to use with a particular domain[7][20]

One of the preferences of anti-phishing deployment is the use of a browser extension [18]. Apozy is a browser extension that uses browser isolation to transform the phishing webpages into read only ones 1 . Another extension is developed by Cyscon Security[10], the tool is an add on that does anti-phishing feed and has a phishing killer, it is targeted for German language users. Table 1 shows a comparison between these two tools.

---

1 https://www.apozy.com

TABLE 1: COMPARISON BETWEEN CYSCON AND APOZY

| Technique | Cyscon | Apozy |
|---|---|---|
| Black list | No | Yes |
| White list | No | No |
| Content | No | Yes |
| User preferences | No | No |

According to Table 1, we find that both tools do not consider user preferences and feedback into their phishing classification and do not use white list as well. We believe that most users are interested in few numbers of trusted websites that require their credentials. Hence having a particular list for each user might better serve the anti-phishing tool.

Based on our extensive survey about anti-phishing techniques and their capabilities and limitations, we found that most of them require extensive calculations and are designed to work in a generic mode and are not customized to a particular user. Therefore, we strongly believe that a user requires a personalized anti-phishing tool or plugin that is customized according to his preferences and does the job efficiently.

According to 2010 Neilson survey [11], the average user visits is 89 websites per month, in a way we can say that the average daily website visits is about 3 websites for an average user. This gives us an indication that the number of private websites that a user visits and submit his credentials to is small and can be managed locally. For this reason we moved toward using personalized white list for each user that maintains trusted domains locally for each user, and if a new URL is requested and the corresponding page requires credentials, then the tool might check that URL against external data repositories such as Phisher man or Phish Tank.

In Section 3, we will explain our Personalized Anti-Phishing Guard (PAPG ), that relies on users' preferences and maintains a works efficiently by maintaining a small whitelist generated using user's feedback. is [7].

### 3. PAPG PLUG-IN

As mentioned in Section 2, a web browser plugin for anti-phishing lack the use of users' preferences to build a robust white list required to check against phishing attacks. In this section, we introduce our PAPG plugin that detects phishing according to predefined rules and benefits from users' preferences to enhance the results.

#### A. PAPG Design

PAPG plug-in works in two phases, the first phase check the requested URL against the white list generated by the user while the second phase perform some processing on the requested URL to decide about phishing, this step is performed if the requested URL was not in the white list. Figure 3 below shows the flow diagram of the PAPG Plugin.

An important process of the PAPG is to extract the domain name from the requested URL, this should be done carefully to avoid any misleading tricks in choosing a domain name or inserting the domain name inside the phishing URL.

#### B. PAPG Implementation

The PAPG plugin was initially developed for Google Chrome using JavaScript and HTML. The design of the icon of the pluging is shown in Figure 4. We used colors to indicate the status of the requested URL. The green color shown in Figure 4.a shows that the requested URL is safe while that in Figure 4.b indicates a phishing URL is not safe.

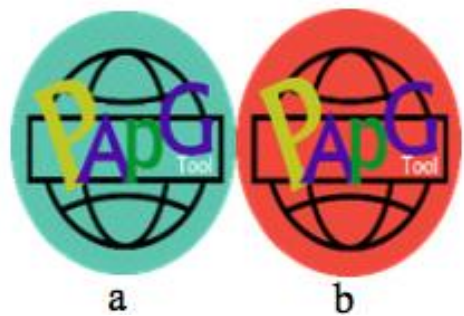

Figure 5: PAPG icon

The result reported by external repositories might be used manually by the user to report that URL as safe or unsafe. If it was reported safe, the user might then add this URL to his whitelist. It is worth mentioning here that this process is carried out rarely for infrequent visited URLs and hence will not degrade the performance of the plugin.

Some enhancements have been added to the PAPG plugin that might better serve users and help them fight phishing, these enhancement will be provided in Section 3.3

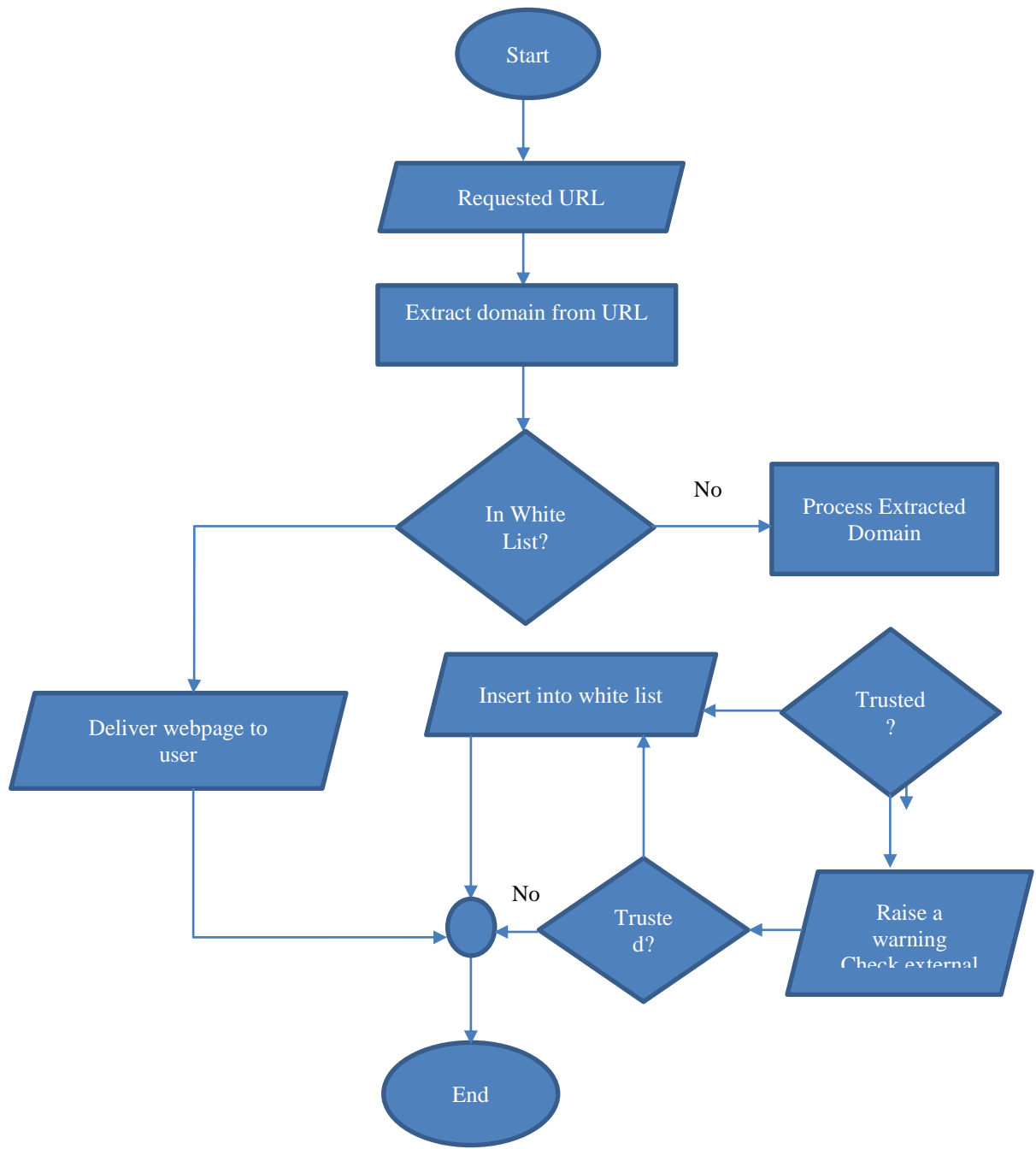

Figure 4: PAPG flowchart

*C. PAPG Design Issues*

To make the plugin acceptable and convenient to use, we enriched it with some features that are required to meet the user's needs. These features are listed below:

*1) Ability to edit the white list*: *The users can adjust the settings of the PAPG plugin including editing the whitelist. Editing includes adding and deleting URLs, as well as setting a configuration password. The configuration password is required to prohibit others from changing the settings of PAPG.*

*2) Automatically add to whitelist*: *PAPG plugin has the ability to configure the white list by automatically adding the frequently visited sites. Before doing so, PAPG prompts the user and update accordingly. This will make it easier for the user to update the whitelist while surfing websites.*

*3) Export and import the whitelist*: *PAPG plugin also enables users to export their whitelist into a file and store it. They are also able to restore that file and use it on another device once needed.*

*4) Automatic delete of unvisited URLs* : *There is a choice for user to delete from white list according to the last day visited. This allows a user to delete whitelist entries never visited last month or 3 months or 6 months as her preferences. This will keep the whitelist small and hence enhance the performance.*

These were the most important issues required by the users who tested PAPG plugin. However, there were also some recommendations that we will consider for future work.

**4. Discussion**

To the best of our knowledge, PAPG is the first personalized anti-phishing plugin that is adapted according to user's feedback. PAPG generates a user's profile that might be migrated from one device to another. The performance of PAPG plugin is discussed below.

According to [11], on average 3 newly websites are visited by a user daily. Hence there will be around 90 per month. Assuming that the user sets PAPG to delete unvisited entries in the white list for 3 months then the list will contain about 270 entries for comparison which is small enough to consider the search time complexity almost constant.

Doing this might in some cases require the toll to perform an extra check for the deleted URL's. However, the probability of using a website that has never been used for 3 months is very small and might be neglected.

The most time consuming process is checking the requested domain against trusted domains using some online repositories. This step requires internet connection and might take some time depending on Internet connection. However, we tried to mitigate the effect by making this process work in the back and during this the user might navigate other websites.

One other issue, is that we tested the plugin among 100 users from Hebron University students, the following discussion analyses their responses to our questionnaire

When students were asked about being victims of a phishing attack 45% answered yes while 20% said no and 35% said that they do not know. This data shows that over one third of students are not sure if they were victims so they need awareness in this field,. Similar results have been reported by [12] in their survey about cybercrime in Palestine and its effects on individuals.

When we asked them about the helpfulness of PAPG and ease of use and whether they recommend it to other users, their responses were as shown in Figures 5,6,7 below.

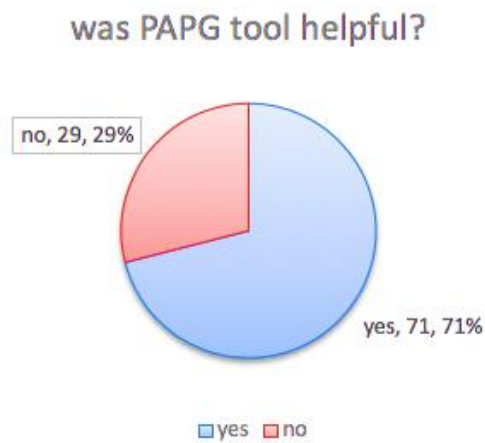

Figure 6: results of whether PAPG was helpful or not In Figure 5, 71% reported that the plugin was helpful to them and in accordance they will be using it. A similar percentage 75% said that the plugin was easy to use and does not need that much training or effort Figure 6. The two figures shows that with some additional efforts of training and awareness, the percentage of acceptance will be enhanced.

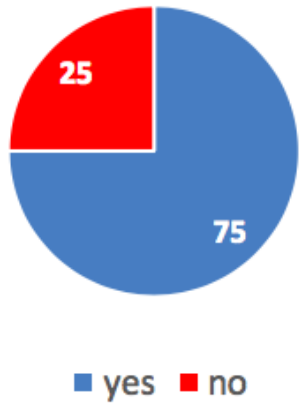

Figure 7: PAPG ease of use

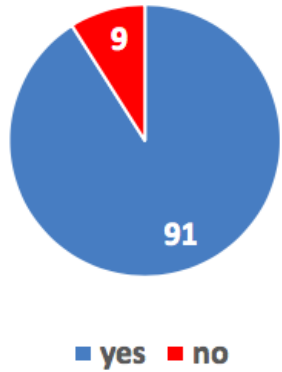

Figure 8: PAPG recommendation

In Figure 7, 91% of students recommend this tool to others, this means that they all care about phishing and that one should take actions to avoid being a victim.

## CONCLUSION

In this paper, we proposed a novel personalized anti-phishing plugin called PAPG. The plugin uses user's preferences to create a robust user profile that will be used as a heuristic for classifying phishing websites. PAPG plugin depends mainly on user's feedback to build a local whitelist for the frequently visited websites. It may also utilize black list repositories over cloud to help decide for domains that are not listed yet in the local white list. The users are able to customize the plugin by editing some fields. For convenience use, the users are able to store and restore their profiles while using different devices from different locations. PAPG works efficiently due to its dependency mostly on a short whitelist that contains trusted domains for a particular user. Different users may have different whitelists according to their preferences. The Plugin was tested by 100 persons who agreed that PAPG is useful and easy to use and they also recommended it for others.

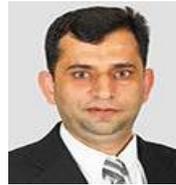

**Belal Amro** is an assistant professor and the head of Computer Science Department at Hebron University - Palestine, where he has been working since 2003. Currently, he is conducting research in network security, wireless security, privacy preserving data mining techniques. From 2003 to 2004, he was a research assistant at Hebron University. From 2005 to 2007, he was an instructor in the Computer Science Department at Hebron University after having his MSc. degree in complexity and its interdisciplinary applications form Pavia- Italy. During 2008-2011 he received an ERASMUS PhD grant in Sabanci University-Turkey. From 2011-2012 he worked as research assistant at Sabanci University. In 2012, Belal received a PhD in Computer Science and Engineering From Sabanci University- Istanbul,, turkey. He he has served as technical program committee member of different international conferences and journals, and reviewed more than 45 paper in the field of information technology including privacy and security.

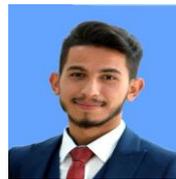

**Ahmad Helal Abusabha** was born in Yatta, Palestine on 20-12-1996, I have done my BSc degree in (Security and Protection of Computer Networks) from Hebron University, Hebron, Palestine, in 2018. Currently I am working on providing security solutions and duties on different freelancing companies. Mr. Ahmad is Interested in information security and investigation of cybercrime.

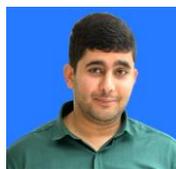

**Ibrahim Ismail Najjar** was born in Dubai, UAE on 7-02-1996, I have done my BSc degree in (Security and Protection of Computer Networks) from Hebron University, Hebron, Palestine, in 2018. Currently I am working on providing security solutions and duties on different freelancing companies. Mr. Ibrahim is Interested in information security and Programming.

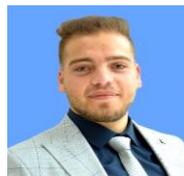

**Ammar Hazim Quneibi** was born in Bethlehem, Palestine on 31-05-1996, I have done my BSc degree in (Security and Protection of Computer Networks) from Hebron University, Hebron, Palestine, in 2018. Currently, I am working on providing security solutions and duties on different freelancing companies. Mr. Ammar is Interested in issues related to cybercrime and network developments.